\documentclass[12pt]{iopart}
\usepackage{iopams}
\usepackage{graphicx}
\usepackage[all]{xy}
\begin{document}
\letter{A new description of spin tunneling in magnetic molecules}
\author{D. Galetti}
\address{Instituto de F\'{\i}sica Te\'{o}rica, Universidade Estadual Paulista, 
         Rua Pamplona 145, 01405-900, S\~{a}o Paulo, SP, Brazil}
\ead{galetti@ift.unesp.br}
\begin{abstract}
A new approach is used that allows to describe the magnetic molecules main
properties in a direct and simple way. Results obtained for the $Fe_{8}$
cluster show good agreement with the experimental data.
\end{abstract}
\pacs{03.65.Ca, 75.45+j, 75.60Jp}
\submitto{\JPC}
\vspace*{1cm}
In recent years, with the experimental advances in the measurement of
magnetic molecular clusters properties, it has emerged a new frontier in
this area. One of the interesting aspects that came out of these studies is
the possibility of measuring spin tunneling in mesoscopic systems what
corresponds, in a standard quantum description, to the tunneling of the
collective degree of freedom corresponding to the magnetization direction
through a potential barrier separating two minima of an effective potential
associated with the spatial orientation\cite
{sessoli,novak,paulsen,friedman,thomas,barra,sangregorio,caneschi}.

From the theoretical point of view, spin tunneling has been treated mainly
by the use of a WKB method adapted to spin systems\cite{vansutto,walter}, by
using Feyman's path integral treatment of quantum mechanics\cite{enz,chud},
and also by using $su(2)$ coherent states\cite{perelomov} in order to
establish a correspondence between the spectrum of the spin system with the
energy levels of a particle moving in an effective potential\cite{ulyanov}.

In the present letter we intend to show that still another approach may be
used for describing spin tunneling -- in angle representation -- in such a
way that analytic expressions for the characteristic parameters of the
magnetic molecular clusters can be obtained; in particular, the results for
the spectrum and energy barrier heights are directly obtained and the spin
tunneling process can be easily interpreted.

The starting point of the present approach is the introduction of a quantum
phenomenological Hamiltonian describing the spin system, written in terms of
angular momentum operators obeying the standard commutation relations, and
that reflects the internal symmetries of the system. It may also contain
terms taking into account external applied magnetic fields. The degree of
freedom that undergoes tunneling is considered a particular collective
manifestation of the system, and it is assumed to be the only relevant one.
At the same time, the temperature of the system is assumed so conveniently
low that possible related termally assisted processes are not taken into
account so that only quantum effects are considered. For instance, the
general quantum Hamiltonian -- sometimes also called giant spin model -- 
\begin{equation}
H=AJ_{z}-DJ_{z}^{2}+\frac{E}{2}\left( J_{+}^{2}+J_{-}^{2}\right) 
\label{eq1}
\end{equation}
can be used to study some systems of interest. In particular, this
Hamiltonian can describe the octanuclear iron cluster\cite{wieg}, $Fe_{8}$,
in the presence of an external magnetic field along the $z$ axis if we take $%
A=g\mu _{B}H_{\parallel }$; by taking $A=0$ no external field is considered.
This spin system has a $j=S=10$ ground state and a suggested pure quantum
spin tunneling below $0.35\;K$; furthermore $D/k_{B}=0.275\;K$, $%
E/k_{B}=0.046\;K$ \cite{barra,caneschi}, and $k_{B}$ is Boltzmann's constant
respectively. On the other hand, from a pure algebraic model point of view,
it is interesting to see that the Lipkin quasi-spin model\cite{lipkin} of
wide use in many-body physics is also obtained by just considering $D=0$. At
the same time that the Hamiltonians in both cases are similar, and in this
form we can compare their results, our interest in this model also resides
in the fact that it stands for a valuable testing ground for checking the
validity of approximations in treating collective degrees of freedom.

In what follows we will show how we can discuss the spin tunneling process
in the $Fe_{8}$ cluster by the use of a new Hamiltonian that is an
approximate version of Eq. (\ref{eq1}). This new Hamiltonian is obtained
through a series of transformations performed on the matrix generated by
calculating the expectation values of Eq. (\ref{eq1}) with the $su(2)$
coherent states $|j,z\rangle $, where $z$ is a complex variable and $j$
characterizes the angular momentum state multiplet\cite{perelomov}. The
Hamiltonian is the overcomplete spin coherent states representation then 
given by  
\begin{equation}
\langle j,z^{\prime }|H|j,z\rangle =K\left( z^{\prime },z\right) ,
\label{eq3}
\end{equation}
the also known generator coordinate energy kernel\cite{griffin}, embodies
the quantum information related to the system we want to study. The
procedure of extracting a new Hamiltonian -- written now in terms of an
angle variable -- from Eq. (\ref{eq3}) has been already shown elsewhere and
will not be repeated here \cite{garu,gapi,ga}. In fact, as it was proved there,  
the variational generator coordinate method can in this case be used to rewrite 
the Hamiltonian from which we start in an exact and discrete representation 
which can then be conveniently treated in order to give an approximate 
Hamiltonian in the angle representation. It is important to point that
the Hamiltonian we obtain is in fact an approximate one, but we have also
shown, by studying the Lipkin model, that it is already a reliable
Hamiltonian for spin systems with $j=S\gtrsim 5$, as it is the case for the $%
Fe_{8}$ cluster, when $S=10$, as mentioned before. Furthermore, since this
approach is based on quantum grounds from the beginning, it does not need to
go through any quantization process. Also, it is not necessary to convert
the discrete spin system into a continuous one as it is usually done\cite
{villain}, at the same time that the quantum character of the angle-angular
momentum pair is properly taken into account.

The Hamiltonian in the angle representation associated with the $Fe_{8}$
cluster in the absence of an external magnetic field is then written
explicitely as 
\begin{equation}
H\left( \phi \right) =-\frac{1}{2}\frac{d}{d\phi }\frac{1}{M\left( \phi
\right) }\frac{d}{d\phi }+V\left( \phi \right) ,  \label{eq4}
\end{equation}
where 
\begin{equation}
V\left( \phi \right) =-\left( D-E\right) S\left( S+1\right) \cos ^{2}\phi
-ES\left( S+1\right)   \label{eq5}
\end{equation}
is the potential energy, whereas the effective ``mass'' is given by 
\begin{equation}
M\left( \phi \right) =\frac{1}{2\left( D-E\right) \cos ^{2}\phi +4E},
\label{eq6}
\end{equation}
with $-\pi <\phi \leq \pi $. It is important to observe that the effective
``mass'' is not constant over the angle domain and that it plays an
essential role in this description. Figure 1 shows the potential function as
well as the effective ``mass''. The minima of both functions occur at $\phi
=0,\pi $ while the maxima occur at $\phi =$ $\frac{\pi }{2},\frac{3\pi }{2}%
(=-\frac{\pi }{2})$, as expected.

The eigenvalues and eigenfunctions associated with the Hamiltonian (\ref{eq4}%
) can then be directly obtained by numerically solving the Schr\"{o}dinger
equation \ 
\begin{equation}
H\left( \phi \right) \psi _{k}\left( \phi \right) ={\cal E}_{k}\psi
_{k}\left( \phi \right)   \label{eq7}
\end{equation}
by means of a Fourier analysis. The ground state energy thus obtained, $%
{\cal E}_{gs}\simeq -27.6447\;K$, agrees quite well with the result obtained
by numerically diagonalizing the exact phenomenological Hamiltonian Eq. (\ref{eq1})
within the $|j=S,m\rangle $ state basis (hereafter these results will be
called the reference values), specifically, the deviation from the reference
value is of the order of $0.5\%$. Figure 2 depicts the wave functions
associated with the lowest pair of energy eigenstates. In what concerns the
energy splitting of those states, we obtain a result that is of the same
order of magnitude as the reference value; in fact, the deviation  is of the
order of $17.5\%$, being that our result is smaller than the reference value.

Now, we see that the top of the potential barrier is explicitly given by 
\begin{equation}
V_{\max }\left( \phi \right) =-ES\left( S+1\right)   \label{eq8}
\end{equation}
so that 
\begin{equation}
h_{b}=-ES\left( S+1\right) -{\cal E}_{gs}  \label{eq9}
\end{equation}
measures the ground state energy barrier. Using our result for the ground
state we get $h_{b}\simeq 22.58\;K$, which is only $1.7\%$ higher than the
experimental result, namely $22.2\;K$, presented in \cite{barra}. An
approximate analytic expression for the ground state energy barrier can be
obtained from another perspective. To this end, we first take into account
that 
\begin{equation}
{\cal E}_{\min }=V\left( \phi =0\right) =-DS\left( S+1\right) =-30.25\;K,\;
\label{eq10}
\end{equation}
and we also assume that the ground state energy is approximately given by $%
{\cal E}_{gs}\simeq {\cal E}_{\min }+\omega /2$. We then perform a harmonic
approximation at the potential minimum, 
\begin{equation}
M\left( \phi \right) \omega ^{2}|_{\phi \left( {\bf E}_{gs}\right) }=\frac{%
d^{2}V\left( \phi \right) }{d\phi ^{2}}|_{\phi _{\min }},\;  \label{eq11}
\end{equation}
from which we obtain 
\begin{equation}
\omega =2\sqrt{\left( D-E\right) \left[ ES\left( S+1\right) +\left| {\cal E}%
_{gs}\right| \right] }.  \label{eq12}
\end{equation}
Taking advantage of this dependence on $\left| {\cal E}_{gs}\right| $ we
obtain the analytic expressions 
\begin{equation}
\fl
\left| {\cal E}_{gs}\right| =DS\left( S+1\right) +\frac{D-E}{2}\left[ 1-%
\sqrt{1+4\frac{D+E}{D-E}S\left( S+1\right) }\right] \simeq 27.52\;K,
\label{eq13}
\end{equation}
and 
\begin{equation}
\fl
h_{b}=\left( D-E\right) S\left( S+1\right) +\frac{D-E}{2}\left[ 1-\sqrt{1+4%
\frac{D+E}{D-E}S\left( S+1\right) }\right] \simeq 22.46\;K\;  \label{eq14}
\end{equation}
for the ground state energy and barrier respectively. The deviation in the
ground state energy is then $0.08\%$ while for the barrier height it is $%
1.17\%$. Even if we consider the crude approximation $\left( D+E\right)
/\left( D-E\right) \sim 1.0$ we get 
\begin{equation}
\left| {\cal E}_{gs}\right| \simeq DS^{2}+ES=27.96\;K,  \label{eq15}
\end{equation}
and 
\begin{equation}
h_{b}\simeq \left( D-E\right) S^{2}=22.90\;K,  \label{eq16}
\end{equation}
respectively; the value for the barrier height still is in good agreement
with the experimental result.

Now, if an external magnetic field paralel to the $z$ axis is applied the
potential function reads 
\begin{equation}
V\left( \phi \right) =-\left( D-E\right) S\left( S+1\right) \cos ^{2}\phi -%
\sqrt{S\left( S+1\right) }g\mu _{B}H_{\parallel }\cos \phi -ES\left(
S+1\right) ,  \label{eq17}
\end{equation}
while the expression for the effective ``mass'' is given by 
\begin{equation}
M\left( \phi \right) =\frac{1}{2\left( D-E\right) \cos ^{2}\phi +\frac{g\mu
_{B}}{S}H_{\parallel }\cos \phi +4E}.  \label{eq18}
\end{equation}
It is immediate to see that the presence of the magnetic field does not
change the position of the potential minima, $\phi =0,\pi $, but it
introduces a shift in energy at these points so that the difference in the
height between the two minima will be $\left( H_{\parallel }>0\right) $%
\begin{equation}
V\left( \phi =\pi \right) -V\left( \phi =0\right) =2\sqrt{S\left( S+1\right) 
}g\mu _{B}H_{\parallel }.  \label{eq19}
\end{equation}
This means that for some particular values of $H_{\parallel }$ there will
occur a degeneracy in the energy spectrum such that the state with, for
instance, $m=n$ will match its energy with that of the state with $m^{\prime
}=-n+k$, as it is indeed expected. The particular value of the magnetic
field increment $H_{0}$ that leads to the matching of the energy levels,
therefore given rise to the appearance of the degeneracies, $H_{\parallel
}=k $ $H_{0}$, can be obtained from an analysis of the effective ``mass''
expression. Realizing that the presence of zeroes in the function $I\left(
\phi \right) =1/M\left( \phi \right) $ (infinities of the effective
``mass'') indicates that tunneling cannot occur, we look for the expression
for the strength of $H_{\parallel }$ beyond which tunneling will not take
place. A direct calculation shows that the limit is given by

\begin{equation}
H_{\parallel }^{\lim }=\frac{4Sk_{B}}{g\mu _{B}}\sqrt{2E\left( D-E\right) }%
\simeq 4.32\;T  \label{eq20}
\end{equation}
so that 
\begin{equation}
H_{0}=\frac{H_{\parallel }^{\lim }}{2S}=\frac{2k_{B}}{g\mu _{B}}\sqrt{%
2E\left( D-E\right) }\simeq 0.216\;T  \label{eq21}
\end{equation}
for $g=2$. This result is in good agreement with the experimental value $%
H_{0}=0.22\;T$\cite{barra,caneschi}. It can be immediately seen that for
this value of the external paralel magnetic field the original minimum at $%
\phi =\pi $ and the maxima at $\phi =\pi /2$ and $\phi =3\pi /2$ have turned
into a single maximum of the potential function, while the only surviving
minimum is the one at $\phi =0\;(=2\pi )$. This means that, in this
particular situation, there is only one direction the spin can be directed
at.

In conclusion we have proposed an angle-based description of the spin
tunneling in magnetic molecules that can account for the basic results
governing this kind of phenomenon. The results obtained in the case of the $%
Fe_{8}$ cluster agree quite well with the experimental data and the
interpretations follow in a direct and simple way.

The author is grateful to Prof. Frederico F. S. Cruz from 
Universidade Federal de Santa Catarina for 
valuable suggestions. The partial support of the Conselho
Nacional de Desenvolvimento Cient\'{i}fico e Tecnol\'{o}gico, CNPq, Brazil,
is gratefully acknowledged.

\newpage

Figure Captions

\bigskip

Figure 1: The potential and effective ``mass'' functions associated with the 
$Fe_{8}$ cluster with parameters $D/k_{B}=0.275\;K$ and $E/k_{B}=0.046\;K$.
The interval is taken as $-\pi /2<\phi \leq 3\pi /2$ for clarity.

\bigskip Figure 2: $Fe_{8}$ cluster ground state wave function $\psi
_{0}\left( \phi \right) $ and first excited state wave function $\psi
_{1}\left( \phi \right) $.

\end{document}